\documentclass[preprint,12pt]{elsarticle}

\usepackage{amssymb}

\usepackage{listings}
\usepackage{hyperref} 
\usepackage{xcolor} 

\newcounter{bla}

\journal{Computer Physics Communications}

\begin{document}

\begin{frontmatter}



\title{\textsc{GenASiS }\texttt{Basics}: Object-oriented utilitarian functionality for large-scale physics simulations (Version~4)}


\author[a]{Reuben D. Budiardja\corref{author}}
\author[b]{Christian Y. Cardall\fnref{copyright}}

\cortext[author] {Corresponding author.\\\textit{E-mail addresses:} cardallcy@ornl.gov (C.Y. Cardall), reubendb@ornl.gov (R.D. Budiardja).}
\address[a]{National Center for Computational Sciences, Oak Ridge National Laboratory, Oak Ridge, TN 37831-6354, USA}
\address[b]{Physics Division, Oak Ridge National Laboratory, Oak Ridge, TN 37831-6354, USA}

\fntext[copyright]{This manuscript has been authored by UT-Battelle, LLC under Contract No. DE-AC05-00OR22725 with the U.S. Department of Energy. The United States Government retains and the publisher, by accepting the article for publication, acknowledges that the United States Government retains a non-exclusive, paid-up, irrevocable, worldwide license to publish or reproduce the published form of this manuscript, or allow others to do so, for United States Government purposes. The Department of Energy will provide public access to these results of federally sponsored research in accordance with the DOE Public Access Plan (http://energy.gov/downloads/doe-public-access-plan).}

%
%
%
%
%
%

\begin{abstract}
\textsc{GenASiS} \texttt{Basics} provides modern Fortran classes furnishing extensible object-oriented utilitarian functionality for large-scale physics simulations on distributed memory supercomputers. 
This functionality includes physical units and constants; display to the screen or standard output device; message passing; I/O to disk; and runtime parameter management and usage statistics.
This revision---Version 4 of \texttt{Basics}---includes a name change and additions to functionality, including the facilitation of direct communication between GPUs.
\end{abstract}


\end{frontmatter}



{\bf NEW VERSION PROGRAM SUMMARY}

\begin{small}
\noindent
{\em Program Title:} \\
\texttt{SineWaveAdvection}, \texttt{SawtoothWaveAdvection}, and \texttt{RiemannProblem} (fluid dynamics example problems illustrating \textsc{GenASiS} \texttt{Basics}); \texttt{ArgonEquilibrium} and \texttt{ClusterFormation} (molecular dynamics example problems illustrating \textsc{GenASiS} \texttt{Basics})\\  
{\em CPC Library link to program files:} (to be added by Technical Editor) \\
{\em Developer's repository link:} \\
https://github.com/GenASiS \\
{\em Code Ocean capsule:} (to be added by Technical Editor)\\
{\em Licensing provisions:} \\
GPLv3  \\
{\em Programming language:} \\
Modern Fortran; OpenMP (tested with recent versions of GNU Compiler Collection (GCC), Cray Compiler Environment (CCE), IBM XL Fortran compiler)                                    \\
{\em Journal reference of previous version:}                  \\
Computer Physics Communications 244 (2019) 483                \\
{\em Does the new version supersede the previous version?:} \\
Yes  \\
{\em Reasons for the new version:}\\
This version includes a significant name change, some minor additions to functionality, and two major additions to functionality: support for systems using AMD GPUs and
infrastructure facilitating GPU-aware MPI communications. \\
{\em Summary of revisions:}


The \texttt{CONSTANT} singleton has been updated to 2022 values \cite{1}.

The class \texttt{MeasuredValueForm}---a class for handling numbers with labels to provide means of dealing with units---has been renamed \texttt{QuantityForm}.

An \texttt{AddCommand} and \texttt{MultiplyAddCommand} have been added to the \texttt{ArrayOperations} division of the code.

The \texttt{Real\_1D\_Form} and \texttt{Real\_3D\_Form} classes, used to construct ``ragged arrays,'' now have \texttt{AllocateDevice ( )} methods to provide mirror allocation of GPU memory.

\texttt{Show\_Command} now has an option to allow the display of more digits for integer and real numbers.

In the \texttt{CurveImageForm} and \texttt{StructuredGridImageForm} classes used for I/O, the \texttt{SetGrid} and \texttt{SetReadAttributes} methods have been replaced by \texttt{SetGridWrite} and \texttt{SetGridRead} respectively. 
An optional flag \texttt{StorageOnlyOption} of their \texttt{Read} methods provides streamlined data input that assumes the data being read conforms to the grid resolution and domain decomposition of the currently running program.  

In the \texttt{PROGRAM\_HEADER} singleton, the method \texttt{RecordStatistics} replaces \texttt{ShowStatistics} for recording memory usage and timers. 
The recording of these data has been refactored and streamlined.
Memory usage statistics are now available on macOS. 
In order to facilitate organic ordering of timer data corresponding the order they are encountered in the code (so as to avoid the necessity of hard-coded timer setup routines), the \texttt{AddTimer} method has been deleted, and the \texttt{Timer} method returns a pointer to either an existing instance of \texttt{TimerForm} or a new one, if it does not yet exist. WARNING: It is important to initialize timer handle variables to zero (or a negative value) in order for the code to recognize that a new timer needs to be created, and to avoid spurious handle values.

GPU-aware MPI communications---passing GPU memory addresses directly to MPI routines---is now supported by the \texttt{MessagePassing} classes (see the original and Version 2 updates of this article for more detailed descriptions of the \texttt{MessagePassing} classes). 
A new method \texttt{AllocateDevice ( )} has been added to these classes to activate this feature. 
When the communication buffers are allocated in an instantiation of the class, \texttt{AllocateDevice ( )} creates a mirror allocation of the buffers on the GPU.
When associations with pre-existing arrays are used as the communication buffers with the class instantiation, \texttt{AllocateDevice ( )} deduces the GPU memory addresses associated with these buffers to be used for future MPI communications.

The example fluid dynamics problem \texttt{RiemannProblem} included in this release has been modified to illustrate the use of GPU-aware MPI.
In the \texttt{DistributedMeshForm} class, a call to the \texttt{AllocateDevice ( )} method is made for the instances of the \texttt{MessageIncoming\_*} and the \texttt{MessageOutgoing\_*} classes when GPU offload is enabled (see \cite{2} and Version 3 of this article for more detailed descriptions of GPU offload in \textsc{GenASiS}).
The use of GPU-aware communication can be explicitly turned on or off using a command-line argument \texttt{DevicesCommunicate=T} or \texttt{DevicesCommunicate=F}, respectively. 
On the Summit supercomputer at the Oak Ridge Leadership Computing Facility (OLCF) \cite{3}, exploiting GPU-aware communications yields over 20\% speedups for \texttt{RiemannProblem} due to the avoidance of explicit GPU-memory to CPU-memory data movement for MPI communications.

The example program \texttt{RiemannProblem} has been modified such that the use of GPU offload can be controlled by a command-line argument \texttt{UseDevice=[T,F]} when the executable is built with OpenMP offload support. 
For example, on OLCF Summit, the following commands build and execute the three-dimensional \texttt{RiemannProblem} with $512^3$ cells three times with eight MPI processes.
The first one, by default, uses GPU offload and GPU-aware communications. 
The second run uses MPI communications on the host.
And finally the third run uses OpenMP threading on the CPU by disabling GPU offload, which also automatically disables GPU-aware communications.
\begin{lstlisting}[frame=tb,numbers=left,numbersep=5pt,xleftmargin=10pt]
export GENASIS_MACHINE=POWER_XL
export OMP_NUM_THREADS=7

make ENABLE_OMP_OFFLOAD=1 RiemannProblem

jsrun -n 8 -g 1 -c 7 --bind packed:8 ./RiemannProblem_POWER_XL \
Dimensionality=3D nCells=512,512,512 

jsrun -n 8 -g 1 -c 7 --bind packed:8 ./RiemannProblem_POWER_XL \
Dimensionality=3D nCells=512,512,512 DevicesCommunicate=F

jsrun -n 8 -g 1 -c 7 --bind packed:8 ./RiemannProblem_POWER_XL \
Dimensionality=3D nCells=512,512,512 UseDevice=F
\end{lstlisting}

Finally, this revision adds support for AMD GPUs and other accelerators supported by the HIP programming model \cite{4} as provided by the new file \texttt{Device\_HIP.c} under the directory \texttt{Modules/Basics/Devices}.
The functionalities provided here are either not currently available in OpenMP or not yet implemented widely, such as inquiry of GPU memory usage and allocation of host page-locked memory. 
The use of \texttt{Device\_CUDA.c} and \texttt{Device\_HIP.c} is mutually exclusive and controlled by the Makefile variables \texttt{DEVICE\_CUDA} and \texttt{DEVICE\_HIP}, respectively. 
An example of how this is done can be found in the machine Makefile \texttt{Makefile\_Cray\_CCE}. 

{\em Nature of problem:} \\
By way of illustrating \textsc{GenASiS} \texttt{Basics} functionality, solve example fluid dynamics and molecular dynamics problems.\\
{\em Solution method:} \\
For fluid dynamics examples, finite-volume. For molecular dynamics examples, leapfrog and velocity-Verlet integration. \\
{\em Additional comments including restrictions and unusual features:}\\
Uses the MPI \cite{5} and Silo \cite{6} libraries.
The example problems named above are not ends in themselves, but serve to illustrate our object-oriented approach and the functionality available though \textsc{GenASiS} \texttt{Basics}. 
In addition to these more substantial examples, we provide individual unit test programs for the individual classes comprised by \textsc{GenASiS} \texttt{Basics}. 

\textsc{GenASiS} \texttt{Basics} is available in the CPC Program Library and also at \\
https://github.com/GenASiS.

\section*{Acknowledgements}

This material is based upon work supported by the U.S. Department of Energy, Office of Science, Office of Nuclear Physics under contract number DE-AC05-00OR22725 and the National Science Foundation under Grant No. 1535130. 
This research used resources of the Oak Ridge Leadership Computing Facility, which is a DOE Office of Science User Facility supported under Contract DE-AC05-00OR22725.


\end{small}






\bibliographystyle{elsarticle-num}
\bibliography{<your-bib-database>}

\begin{thebibliography}{0}

\bibitem{1}R.L. Workman et al. (Particle Data Group), to be published in Prog. Theor. Exp. Phys. 2022, 083C01 (2022)

\bibitem{2}R.D. Budiardja and C.Y. Cardall, Parallel Computing 88, 102544 (2019)

\bibitem{3}https://docs.olcf.ornl.gov/systems/summit\_user\_guide.html

\bibitem{4}https://rocmdocs.amd.com/en/latest/Programming\_Guides/Programming-Guides.html

\bibitem{5}https://www.mpi-forum.org 

\bibitem{6}https://wci.llnl.gov/simulation/computer-codes/silo         
\end{thebibliography}







\end{document}